\documentclass{pasj00}

\begin{document}
\SetRunningHead{Totani}{Size Evolution of Early-Type Galaxies}
\Published{(Submitted on August 23, 2009)}

\title{Size Evolution of Early-Type Galaxies and Massive Compact Objects \\ 
as the Dark Matter}

\author{Tomonori \textsc{Totani} }
\affil{Department of Astronomy, Graduate School of Science, 
Kyoto University, Sakyo-ku, Kyoto 606-8502}
\email{totani@kusastro.kyoto-u.ac.jp}



\maketitle

\begin{abstract}
  The dramatic size evolution of early-type galaxies from $z \sim 2$
  to 0 poses a new challenge in the theory of galaxy formation, which
  may not be explained by the standard picture.  It is shown here that
  the size evolution can be explained if the non-baryonic cold dark
  matter is composed of compact objects having a mass scale of $\sim
  10^5 M_\odot$.  This form of dark matter is consistent with or only
  weakly constrained by the currently available observations.  The
  kinetic energy of the dark compact objects is transferred to stars
  by dynamical friction, and stars around the effective radius are
  pushed out to larger radii, resulting in a pure size evolution.
  This scenario has several good properties to explain the
  observations, including the ubiquitous nature of size evolution and
  faster disappearance of higher density galaxies.
\end{abstract}

\section{Introduction}

It is well established that early-type galaxies have mostly been built
up before redshift $z \sim 0.8$ (e.g., Cirasuolo et al. 2007; Pozzeti
et al. 2007). However, recent observations have revealed that
passively evolving, early-type galaxies at higher redshifts ($z \sim$
1--2) are surprisingly much more compact than the likely present-day
descendants of the same stellar mass (e.g., Longhetti et al. 2007;
Trujillo et al. 2007; Cimatti et al. 2008; Damjanov et al. 2009,
hereafter D09).  Such compact galaxies are not found in the local
universe, and they must disappear (decrease in comoving number density
by a factor $>$5000, Taylor et al. 2009) by structural evolution,
meaning that the evolution must occur ubiquitously, rather than
stochastically, for all such galaxies.  In the plane of $M_*$-$r_e$
(total stellar mass and the effective radius including half of the
projected light), they must be ``puffed up'' with a typical size
increase by a factor of 2--3 (a factor of more than 10 in stellar
density) to move into the region populated by the present-day
galaxies, without significant increase in stellar mass
(Fig. \ref{fig:M-rho-r}).  Furthermore, this evolutionary process
seems to work in a broad context of galaxy formation spanning wide
ranges of mass, radius, and redshifts (D09).

This size evolution problem thus presents a new major challenge for
the theory of galaxy formation and evolution.  Several scenarios have
been proposed, but none of them has convincingly explained it.  Here,
a new scenario is proposed that may contribute to solve the
problem. It is assumed that the dominant component of the non-baryonic
cold dark matter in the universe is dark compact objects (DCOs), like
primordial black holes (PBHs) or substructures in dark haloes, having
a mass scale of $M_\bullet \sim 10^5 M_\odot$. This possibility is not
yet excluded by currently available observations (see \S
\ref{section:DCO-obs}).

In \S\ref{section:other-scenarios}, previously proposed scenarios for
this problem and their potential difficulties are briefly reviewed.
The new scenario of this work will be described in
\S\ref{section:df}, followed by some discussions in 
\S\ref{section:DCO-obs} and \S\ref{section:discussion}.

\section{Proposed Scenarios for the Size Evolution}
\label{section:other-scenarios}

Dissipationless ``dry'' mergers of old stellar systems can increase
galaxy sizes without new star formation activity, but stellar mass
should also be increased.  Bezanson et al. (2009) argued that mass
increase with a scaling of $M_* \propto r_e$, which is expected in the
case of major or equal-mass mergers, cannot solve the problem because
it would violate the constraint from the stellar mass found in
present-day early-type galaxies.

\begin{figure*}
  \begin{center}
    \includegraphics[width=70mm,angle=-90]{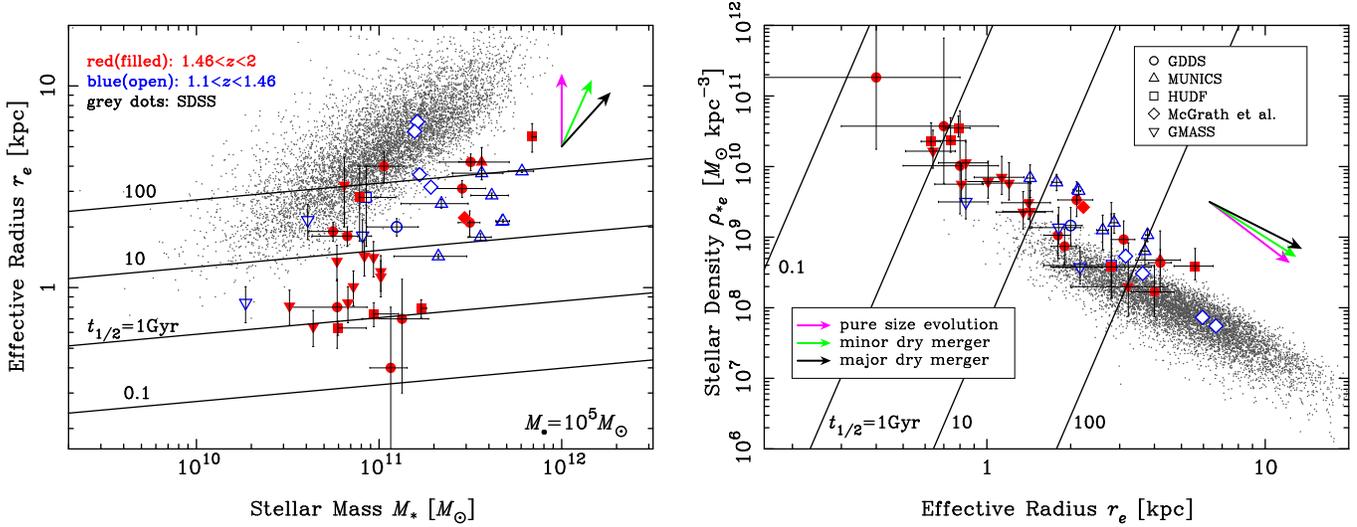} 
  \end{center}
  \caption{The effective radius ($r_e$) versus total stellar mass
    ($M_*$) and stellar density ($\rho_{*e}$, the mean within $r_e$)
    of early-type galaxies.  The red (filled) and blue (open) symbols
    are for high-redshift galaxies ($1.46 < z < 2$ and $1.1 < z <
    1.46$, respectively), derived or complied by D09.  Small grey dots
    are mass estimates by D09 for the local SDSS early-type galaxy
    sample.  All the stellar mass data are corrected to a common
    stellar initial mass function (IMF) by D09.  The contours of the
    energy transfer time scale $t_{1/2}$ are depicted assuming
    $M_\bullet = 10^5 M_\odot$ for DCOs in the both panels ($t_{1/2}
    \propto M_\bullet^{-1}$).  The arrows indicate the evolutionary
    directions expected by a pure-size evolution (magenta), ideal
    minor mergers (green, $r_e \propto M_*^2$), and major (equal-mass)
    mergers (black, $r_e \propto M_*$). }
  \label{fig:M-rho-r}
\end{figure*}

Minor dry mergers ($r_e \propto M_*^2$ in the limit of very large mass
ratio) may be more helpful to solve the problem, as demonstrated by a
simulation of Naab et al. (2009). However, their simulation is only
for one galaxy, and it must be further studied whether this scenario
can explain the size evolution in the general context of galaxy
formation (D09; Williams et al. 2009). All high-$z$ compact galaxies
must experience many minor mergers to gain substantial mass and size
increase, and they would have to be all dry to keep the old stellar
population, which seems somewhat contrived.  It should be noted that
the largest mass ($\sim 10^{12} M_\odot$) of $z \gtrsim 1$ galaxies in
the D09 sample is already comparable with that of the local SDSS
sample (see Fig.  \ref{fig:M-rho-r}), and further mass increase may
contradict the data even in the case of minor mergers.  Theoretically,
minor dry mergers should be effective for massive and high redshift
galaxies (Naab et al. 2009; Hopkins et al. 2009a), but the D09 data
indicate that the required degree of size evolution is similar for
galaxies in the mass range of $M_* \sim 10^{10.5}$--$10^{12} M_\odot$,
and lower mass galaxies disappear faster at higher redshifts
(Fig. \ref{fig:M-rho-r}).  Nipoti et al. (2009) have argued that
constraints from gravitational lensing do not allow size
evolution by more than a factor $\sim$1.8 by dry mergers, which seems too
small to completely solve the size evolution problem.

Puff up of galaxies by mass loss is another possibility, such as
stellar mass loss (D09) or by quasar feedback (Fan et al. 2008).
However, stellar mass loss is not satisfactory because of the small
amount of mass loss and short time scales (D09).  The physical process
of quasar feedback is rather uncertain, and it seems difficult to puff
up quiescent galaxies in passive evolution phase after they lost most
of interstellar gas (Bezanson et al. 2009).  It may require a
significant fine-tuning to explain the ubiquitous size evolution by
rather stochastic quasar feedback processes.

\section{Size Evolution by DCOs}
\label{section:df}

Various observations of nearby early-type galaxies based on stellar
kinetics (e.g., Cappellari et al. 2006), X-ray emitting interstellar
gas (e.g., Fukazawa et al. 2006), and strong/weak gravitational
lensing (e.g., Gavazzi et al. 2007) have shown that stellar mass is
dominant at inner radii while dark matter dominates at outer radii,
and the transition occurs at around the effective radius.  Therefore,
if there is significant transfer of kinetic energy from dark matter to
stars, it would puff up stars at $r \gtrsim r_e$ resulting in a
pure-size evolution of the stellar component.

This energy transfer occurs by dynamical friction.  When stars
have a stellar mass density $\rho_{*e}$ and the Maxwellian velocity
distribution with one-dimensional dispersion $\sigma_{*e}$, the
dynamical friction timescale for a DCO having a velocity $V$ is given
as (Binney \& Tremaine 1987):
\begin{eqnarray}
  t_{\rm df} &\equiv&  -\frac{V}{\dot{V}} 
  = \frac{V^3}{4 \pi G^2 \ln \Lambda \,
    \rho_{*e} M_\bullet \, B(X)}   \ ,
\end{eqnarray}
where the function $B(X) \equiv {\rm erf}(X) - 2X e^{-X^2} /
\sqrt{\pi}$ and $X \equiv V/(\sqrt{2} \, \sigma_{*e})$.  
The Coulomb logarithm $\ln \Lambda = \ln [b_{\max}
V^2/(GM_\bullet)]$ becomes 13.74 and 7.60 if we take the maximum
impact parameter $b_{\max}$ to be 10 kpc (the galaxy scale) and
$(\rho_{*e}/M_\bullet)^{-1/3}$ (the mean separation of DCOs),
respectively, for typical values of $M_\bullet = 10^5 M_\odot$,
$V=200$ km/s, and $\rho_{*e} = 10^{10} M_\odot \rm kpc^{-3}$.  In the
following $\ln \Lambda = 10$ is adopted.

Since $t_{\rm df}$ is strongly dependent on $V$, we should carefully
take the effective mean about $V$. The total kinetic energy $E \propto
\int (M_\bullet V^2/2) f(V)d^3V$ of DCOs will decrease by a factor of
1/2 in a time of $t_{1/2} \equiv E/(2\dot{E})$, where $f(V)$ is the
phase space distribution function of DCOs. Assuming that the initial
velocity distribution of DCOs is the same as that of stars, $t_{1/2}$
is found as:
\begin{eqnarray}
t_{1/2} &=& \frac{B(\sqrt{3/2}) \Gamma(5/2) }{3\sqrt{6} 
\int_0^\infty B(\sqrt{x}) e^{-x} dx} \ t_{\rm av} 
\sim 0.31 \, t_{\rm av}  \ ,
\end{eqnarray}
where $t_{\rm av}$ is $t_{\rm df}$ when $V$ is
the mean value $\langle V^2 \rangle^{1/2} = \sqrt{3} \, \sigma_{*e}$.  

To compare with the observed data in Fig. \ref{fig:M-rho-r}, we
convert $\sigma_{*e}$ into stellar mass using the direct correlation
between the two (i.e., equivalent to the Faber-Jackson relation for a
constant stellar mass-to-light ratio).  The virial relation
$\sigma_{*e}^2 \sim GM_*/r_e$ is another option of this conversion,
but we take the former because the $M_*$-$\sigma_{*e}$ correlation is
tight and the latter could be affected by the size evolution (as
discussed here) and contribution from dark matter.  We expect rather
mild evolution of $\sigma_{*e}$ against the size evolution,
considering the roughly flat velocity profiles of elliptical galaxies
including dark matter (e.g., Mamon \& {\L}okas 2005).  Using observed
velocity dispersions\footnote{The Bernardi et al.'s $\sigma_{\rm obs}$
  has been corrected to an aperture radius of $r_e/8$, and the central
  velocity dispersion is typically $\sim1.4$ times higher than that
  around $r_e$ (e.g., Cappellari et al. 2006). Therefore $\sigma_{\rm
    obs} = 1.4 \, \sigma_{*e}$ has been adopted here.}  $\sigma_{\rm
  obs}$ of the SDSS early-type galaxies (Bernardi et al.  2003) and
the stellar mass estimates of D09, $M_* = 4.7 \cdot 10^{11}
(\sigma_{*e}/{\rm 200 \, km \, s^{-1}})^{3.98} M_\odot$ is found, and
hence the energy transfer time scale becomes
\begin{eqnarray}
\frac{t_{1/2}}{\rm Gyr}
&=&  2.8 \left[\frac{M_*}{10^{11} M_\odot}\right]^{-0.247}
  \left[\frac{r_e}{1 \rm \ kpc}\right]^3 
  \left[\frac{M_\bullet}{10^5 M_\odot}\right]^{-1} .
\end{eqnarray}
Here, the stellar mass fraction of 0.42 within $r_e$ (corresponding to
the S\'ersic index $n=4$, e.g., Mamon \& {\L}okas 2005) has been used
to convert $\rho_{*e}$ (the mean density within $r_e$) into $M_*$ and
$r_e$.

This time scale is compared with the observed data of early-type
galaxies in the plane of $M_*$-$r_e$ as well as $\rho_{*e}$-$r_e$
(Fig. \ref{fig:M-rho-r}).  The high density galaxies ($\rho_{*e}
\gtrsim 10^{10} M_\odot \rm kpc^{-3}$) of the D09 sample is found only
in the higher redshift bin of $z > 1.46$ (red symbols in
Fig. \ref{fig:M-rho-r}), indicating that these galaxies should
disappear on a shorter time scale of a few Gyr, while galaxies of
lower densities ($\rho_{*e} \lesssim 10^{10} M_\odot \rm kpc^{-3}$,
blue symbols) seem to disappear more slowly on a time scale
of $\sim 10$ Gyr.  Fig. 10 of D09 has shown this trend
even more clearly in a $\rho_{*e}$-$z$ plot, where galaxies having
higher stellar density disappear at higher redshifts.  This trend
is expected in the proposed scenario, because galaxies having higher
stellar density are located in the region of shorter $t_{1/2}$, as
indicated in Fig. \ref{fig:M-rho-r}.

The $t_{1/2}$ of high-$z$ early-type galaxies with the largest mass is
larger than the age of the universe.  We may need a larger DCO mass
than the assumed $10^5 M_\odot$, or some other mechanisms, for these
galaxies to evolve into the location of the local galaxies.  On the
other hand, Mancini et al. (2009) has recently reported that the sizes
of the most massive early-type galaxies at $z \sim 1.5$ are similar to
those of local galaxies, indicating a possibility that the size shift
in the $M_*$-$r_e$ plane is smaller for more massive galaxies. In this case
DCOs with $M_\bullet \sim 10^5 M_\odot$ could be sufficient to solve
the problem.

It should be noted that stellar density is larger than that of dark
matter at $r \ll r_e$ (e.g., Mamon \& {\L}okas 2005), and we do not
expect a strong structural evolution of stars at $r \ll
r_e$.\footnote{The accumulation of DCOs in a galactic center may
  affect the innermost stellar distribution, such as stellar core
  formation, but it would also be complicated by the existence of
  super-massive black holes. This is an interesting issue to be
  investigated, but beyond the scope of this letter.}  This is in
contrast to the change of stellar distribution by stellar
remnant black holes that have the same initial density profile as that
of stars (e.g., Merritt et al. 2004).  The small effect at $r \lesssim
r_e$ is consistent with the observations indicating that the core
densities of high-$z$ elliptical galaxies are not much different from
those of local galaxies (Bezanson et al. 2009; Hopkins et al. 2009b).
The expected mild evolution of velocity dispersion is also consistent
with the observational result of Cenarro \& Trujillo (2009).\footnote{
  But see also van Dokkum et al. (2009) who have reported a large
  velocity dispersion for one galaxy at $z=2.186$.  Velocity
  dispersion measurements of early-type galaxies at $z \gtrsim 1$ are
  still a difficult task, and more observational studies are highly
  desired.}

To conclude, the proposed scenario has several features that are good
to explain the observed trends: (1) the evolution should be pure-size
evolution, (2) it works ubiquitously for all galaxies in a
non-stochastic manner on cosmological time scales, (3) higher density
galaxies should disappear on shorter time scales, (4) stellar density
at $r \ll r_e$ is not much affected, and (5) evolution of velocity
dispersion should be mild.

\section{Observational Constraints on DCOs}
\label{section:DCO-obs}

There are no strong observational constraints to exclude the
possibility of $10^5 M_\odot$ DCOs as the dominant form
of dark matter (for reviews see Mack et al. 2007; Ricotti et al. 2008,
hereafter M07 and R08).  DCOs of $\lesssim 10^5 M_\odot$ are
marginally allowed by constraints from strong lensing of compact radio
galaxies (Wilkinson et al. 2001) and millilensing of gamma-ray bursts
(Nemiroff et al. 2001). On the other hand, anomalous flux ratios in
gravitational lenses may be explained by DCOs of $10^{5-6} M_\odot$
rather than dark matter of particle-mass scale (Mao et al. 2004).

Yoo et al. (2004) declared ``the end of MACHO era'' by closing the
window of $M_\bullet \sim 30$--$10^6 M_\odot$ based on the existence
of wide binary stars in our Galaxy halo, but the robustness of their
conclusion has been questioned by Quinn et al. (2009) using new data.

DCOs of $M_\bullet \lesssim 10^5 M_\odot$ do not affect the stellar
distribution of dwarf spheroidal galaxies (dSphs) like Draco in the
Milky Way halo (Jin et al. 2005), and $10^{5-6} M_\odot$ DCOs may even
be helpful to solve some problems in galaxy formation such as disk
heating (Lacey \& Ostriker 1985) and the core/cusp problem (Jin et
al. 2005).  S\'anchez-Salcedo \& Lora (2007) has claimed that $10^5
M_\odot$ DCOs are excluded as dark matter from the survival of the
kinetically cold stellar clump found in the Ursa Minor (Kleyna et
al. 2003, hereafter K03).  However, quantitative constraint on the
velocity dispersion of the cold clump was not derived by K03, making a
quantitative analysis difficult. The origin and history of the clump
is also still uncertain (see, e.g., Mu\~noz et al. 2008; Penarrubia et
al. 2009).  It should be noted that UMi is showing substantial
morphological distortions, unlike most other dSphs (K03).

Afshordi et al. (2003) excluded DCOs heavier than a few times $10^4
M_\odot$ as dark matter, using the small-scale density fluctuations
inferred by Ly$\alpha$ forests.  However, this analysis is depending
on a modeling of complicated baryonic physics about Ly$\alpha$ forests
that may include significant systematic uncertainties.

Mii \& Totani (2005) and Mapelli et al. (2006) discussed the expected
number of X-ray sources in nearby galaxies considering radiatively
inefficient accretion flow onto halo black holes of $M_\bullet \gtrsim
10^2 M_\odot$. R08 set strong constraints on PBHs as dark matter by
requiring that the accretion activity onto PBHs does not affect the
anisotropy of the cosmic microwave background.  However, these
calculations include many assumptions and large uncertainties
especially about accretion physics and feedback effects. R08 discussed
about angular momentum estimating it from the normal cosmic density
fluctuation. However, the proper mean separation of $10^5 M_\odot$
dark matter DCOs at $z \sim 1000$ is about 10 pc, which is comparable
to the Bondi radius at that epoch estimated by R08.  Then interaction
with nearby DCOs would result in significant angular momentum that
should prevent efficient accretion.

\section{Discussion}
\label{section:discussion}

Although DCOs are less popular than elementary particles as the
candidate of dark matter, several formation scenarios have been
proposed (see M07 and R08 for reviews).  PBHs can be formed by density
fluctuation of a modest amplitude or softening of equation of state,
when the fluctuation scale enters the event horizon of the universe.
The mass scale of $10^5 M_\odot$ corresponds to the horizon mass
($M_{\rm hor} = tc^3/G$) at a temperature of $\sim$ MeV, i.e., close
to the $e^{\pm}$ annihilation era.  Note that only a small portion
($\sim 10^{-6}$) of the universe should be converted into PBHs to
explain the dark matter density, and hence it may not necessarily
affect the primordial nucleosynthesis.

DCOs need not to be very compact like PBHs, but substructures in dark
haloes generally found in simulations of structure formation should
also contribute to dynamical heating of stars.  However, the mass
fraction of substructure must be of order unity to explain the size
evolution, which seems rather unlikely within the standard picture
because less than $\sim$10\% of dark mass within the virial radius
(and even smaller fraction at inner radii) is found in substructures in recent
simulations (Springel et al. 2008).

The black hole masses found in active galactic nuclei are always
$\gtrsim 10^6 M_\odot$. Black holes heavier than $10^9 M_\odot$ have
been discovered in quasars beyond $z > 6$ (e.g., Willott et al. 2003),
and their existence at such high redshifts is not easy to explain by
evolution starting from stellar mass black holes (e.g., Tanaka \&
Haiman 2009).  It is obvious that the explanation would become easier,
if dark matter is composed of $10^5 M_\odot$ DCOs.

More realistic, numerical studies of galaxy size evolution would be
required to confirm this hypothesis. Observational searches for DCOs
by various approaches (as discussed in \S \ref{section:DCO-obs}) may
detect DCOs in the future.

\medskip

The author would like to thank I. Damjanov for kindly providing their
data used in D09.  This work was supported by the Grant-in-Aid for the
Global COE Program ``The Next Generation of Physics, Spun from
Universality and Emergence'' from the Ministry of Education, Culture,
Sports, Science and Technology (MEXT).


\end{document}